\documentclass[a4paper]{jpconf}
\usepackage{graphicx}
\usepackage{amsmath}
\def\ll#1#2{\tilde{\lambda}_{#1}.\tilde{\lambda}_{#2}}
\begin{document}
\title{Basics of doubly heavy tetraquarks}

\author{A Valcarce$^1$, JM Richard$^2$ and J Vijande$^3$}

\address{$^1$ Departamento de F\'\i sica Fundamental e IUFFyM, 
           Universidad de Salamanca, E-37008 Salamanca, Spain}

\address{$^2$ Universit\'e de Lyon, Institut de Physique Nucl\'eaire de Lyon,
           IN2P3-CNRS--UCBL,\\ 4 rue Enrico Fermi, 69622  Villeurbanne, France}

\address{$^3$ Unidad Mixta de Investigaci\'on en Radiof\'\i sica e Instrumentaci\'on Nuclear en Medicina (IRIMED),
           Instituto de Investigaci\'on Sanitaria La Fe (IIS-La Fe),
           Universitat de Valencia (UV) and IFIC (UV-CSIC), Valencia, Spain}

\ead{valcarce@usal.es, j-m.richard@ipnl.in2p3.fr, javier.vijande@uv.es }

\begin{abstract}
We outline the most important results regarding the stability of doubly heavy tetraquarks $QQ\bar q\bar q$
with an adequate treatment of the four-body dynamics. We consider both color-mixing and spin-dependent effects. 
Our results are straightforwardly applied to the case of all-heavy tetraquarks $QQ\bar Q\bar Q$. We conclude
that the stability is favored in the limit $M_Q/m_q \gg 1$ pointing to the stability of the $bb\bar u\bar d$ state
and the instability of all-heavy tetraquarks. 
\end{abstract}

\section{Introduction}
Despite the impression given by the recent flurry of studies dealing with multiquark states,
flavor-exotic multiquarks have already a long history~\cite{Ade82} and have motivated an abundant literature
(see Ref.~\cite{Ric18} for a recent compendium). In the pioneering 
work of Ader, Richard and Taxil~\cite{Ade82} it was shown that
$QQ\bar q \bar q$ four-quark configurations become more and more
bound when the mass ratio $M_Q/m_q$ increases. The critical value 
of $M_Q/m_q$ for binding is somewhat model dependent.

Currently, a broad theoretical consensus about the existence of a stable
axial vector doubly bottom tetraquark has been reached.
Lattice QCD calculations find unambiguous signals for a stable $J^P=1^+$ 
bottom-light tetraquark~\cite{Fra17}. Based on a diquark hypothesis, Ref.~\cite{Kar17} 
uses the discovery of the $\Xi_{cc}^{++}$ baryon to {\em calibrate} the binding energy in a $QQ$ diquark. Assuming that
the same relation is true for the $bb$ binding energy in a tetraquark,
it concludes that the axial vector $bb\bar u \bar d$ state is stable.
The Heavy-Quark Symmetry analysis of Ref.~\cite{Eic17} predicts the existence
of narrow doubly heavy tetraquarks. Using as input for the doubly bottom baryons, not 
yet experimentally measured, the diquark-model calculations
of Ref.~\cite{Kar17} also leads to a bound axial vector $bb\bar u \bar d$ 
tetraquark. Other approaches, using Wilson twisted mass lattice QCD~\cite{Bic16}, 
also find a bound state. Few-body calculations using quark-quark 
Cornell-like interactions~\cite{Vij09},
simple color magnetic models~\cite{Luo17}, QCD sum rule analysis~\cite{Duc13}, or
phenomenological studies~\cite{Cza18} come to similar conclusions. More doubtful has become
the prediction about the stability of all-heavy tetraquarks~\cite{Hug18}.

In the present note, we stress that a careful treatment of the few-body problem 
is required before drawing any conclusion about the existence of stable states in 
a particular model. There is, indeed, a dramatic spread of strategies: some 
authors use the full machinery of a variational method based on correlated 
Gaussians or hyperspherical expansion, and others use a crude trial wave function 
or a cluster approximation.

\section{General results based on symmetry breaking}
\label{sec-sb}
The analogy between the stability of few-charge systems
and multiquarks in additive chromoelectric potentials
offers a good guidance for identifying the favorable
configurations. There are, however, some differences
mainly due to the color algebra replacing the simpler
algebra of electric charges.
Unlike in the case of the positronium molecule, the equal-mass
tetraquarks are unstable in the chromoelectric model
with frozen color wave functions~\cite{Ric17,Val18}. 
In both the atom and quark cases, the four-body system and its threshold, 
after simple rescaling, are governed by a generic Hamiltonian

\begin{equation}
 \label{eq:H4}
 H=\sum_i \frac{{\vec p}_i^{\,2}}{2\,m_i}-{\rm c.o.m.} + \sum_{i<j} g_{ij}\,v(r_{ij}) \, , \;\;\; \sum_{i<j} g_{ij}=2~,
\end{equation}
with $v(r)=-1/r$ in the atomic case, and $v(r)=-a/r+b\,r$ in the quark case~\cite{Eic75}.

In quantum mechanics, the minimum of a Hamiltonian containing a symmetric and an antisymmetric 
term is always lower than the minimum of  the symmetric part. From this result, one can analyze the effect 
of symmetry breaking in systems of four-charged particles. Let us first consider the hydrogen molecule, $M^+ M^+ m^- m^-$.
The Hamiltonian for this system reads,
\begin{equation}
\begin{aligned}
H & = \frac{\vec p_1^{\,2}}{2 \, M} +
\frac{\vec p_2^{\,2}}{2 \, M} +
\frac{\vec p_3^{\,2}}{2 \, m} +
\frac{\vec p_4^{\,2}}{2 \, m} + V = H_0 + H_1 \\
&= \left[\sum_{i}\frac{\vec p_i^{\, 2}}{2 \, \mu} \, + \, V \right] \, + \, \left(\frac{1}{4\,M} \, - \, \frac{1}{4\,m} \right)\left(
\vec p_1^{\,2} + \vec p_2^{\,2} - \vec p_3^{\,2} - \vec p_4^{\,2} \right) \, ,
\end{aligned}
\label{HCP}
\end{equation}
where $2\,\mu^{-1}=M^{-1}+m^{-1}$. The $C$-parity breaking term, $H_1$, lowers the ground state
energy of $H$ with respect to the $C$-parity even part, $H_0$, which is simply a rescaled version of
the Hamiltonian of the positronium molecule. Since $H_0$ and $H$
have the same threshold, and since the positronium molecule is stable, the hydrogen molecule
is even more stable, and stability improves when $M/m$ increases. 
Clearly, the Coulomb character of $V$ hardly matters in this reasoning.  
The key property is that the potential does not change when the masses are modified. 

One can use the same reasoning to study the stability of four-charged particles 
when $C$-parity is preserved but particle symmetry is broken, in other words the
$M^+ m^+ M^- m^-$ configuration. The Hamiltonian is that of Eq.~(\ref{HCP}) by 
exchanging $2\leftrightarrow 3$.
The same arguments used above lead to the conclusion that this configuration gains 
binding with respect to the threshold $(M^+m^-)+(M^-m^+)$ that it shares with $H_0$.
However, there is another threshold that lies lower, $(M^+M^-)+(m^+m^-)$. This 
threshold gains more from the symmetry breaking than the four-body molecule, and, 
indeed, it is found  that the molecule becomes unstable for $M/m \geq 2.2$. 

\begin{figure}[t]
\parbox[b]{.49\textwidth}{%
\includegraphics[width=.45\textwidth]{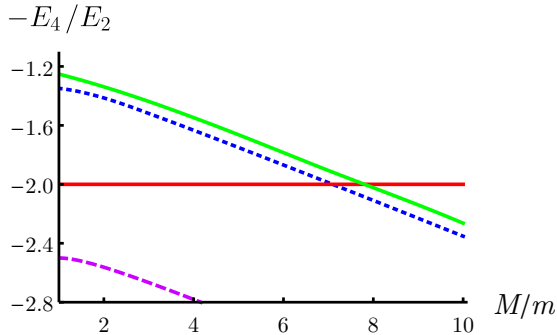}}
\parbox[b]{.49\textwidth}{%
\caption{Comparison of the variational upper bound (light grey solid curve) and Hall-Post lower bound (dotted dark curve) 
for the tetraquark Hamiltonian~(\ref{eq:H4}) with a potential $-r_{ij}^{-1}$. Also shown is the naive 
diquark-antidiquark approximation (dashed faint curve). The faint solid line at $-E_4/E_2=-2$ stands for 
the threshold. Figures are colored online.}}
\label{fig:HPC}
\end{figure}

The above arguments can be directly translated to four-quark systems:
the $QQ\bar q \bar q$ configuration  becomes more and more
bound when the mass ratio $M_Q/m_q$ increases. This has been first established 
in Ref.~\cite{Ade82}, and discussed and confirmed in further studies. 
Arguments based on diquarks, as e.g.~\cite{Kar17}, might considerably 
overestimate the binding, as analyzed in~\cite{Ric18}. 
There are  many variants of the so-called diquark model. An extreme point of view is that 
diquarks are almost-elementary objects, with their specific interaction with quarks and 
between them. In the case of doubly heavy baryons $QQq$ there is obviously a $QQ$ clustering 
which makes it tempting to use a two-step approach:  first a $(QQ)$ diquark and then a 
$(QQ)q$ quasi-meson, as the diquark has the same color $\bar 3$ as an antiquark. 
The exercise can be repeated for the $QQ\bar q\bar q$ states. For simplicity, we consider 
only the case of a frozen $\bar 33$ color wave function, i.e., the Hamiltonian~(\ref{eq:H4}). 
In Fig.~\ref{fig:HPC}, we compare the exact solution 
of~(\ref{eq:H4}) with the approximation consisting of first computing the $QQ$ diquark with 
$r_{12}/2$ alone and $qq$ with $r_{34}$ alone, and then $(QQ)(\bar q\bar q)$ as a meson with a 
potential $r_{12,34}$ and constituent masses $2\,M$ and $2\,m$. 

A remaining problem is to understand why the positronium
molecule lies slightly below its dissociation threshold, while a chromoelectric
model associated with the color additive rule does not bind (at least according 
to most computations). This is due to a 
larger disorder in the color coefficients than in the electrostatic
strength factors entering the Coulomb potential~\cite{Val18}.  
An alternative proof is based on the so-called Hall-Post inequalities~\cite{Hal67,Bas90}.
The principle is rather simple. If a Hamiltonian is decomposed as a sum of Hamiltonians,
\begin{equation}
H= h' + h'' + \cdots \, ,
\end{equation}
then for the lowest energy,
\begin{equation}
E(H) \ge E(h') + E(h'') + \cdots \, .
\label{qq}
\end{equation}
With a $\bar 3 3$ color wave function and a quark mass set to $M=1/2$ for simplicity, 
the Hamiltonian of the all-heavy tetraquark $QQ\bar Q \bar Q$ can be written as~\cite{Val18},
\begin{equation}
H_4=\sum_{i}{\vec p_i^{\,2}} \, + \, \frac{1}{2} \left( V_{12} + V_{34} \right) + \frac{1}{4} \left( V_{13} + V_{14} + V_{23} + V_{24} \right) \, ,
\end{equation}
where $V_{ij}=v(r_{ij})$ is the quarkonium potential. Now, we can rewrite this expression as,
\begin{equation}
H_4=\frac{1}{2} \left( h_{12} + h_{34} \right) + \frac{1}{4} \left( h_{13} + h_{14} + h_{23} + h_{24} \right) \, ,
\end{equation}
where $h_{ij}= \vec p_i^{\, 2} + \vec p_j^{\, 2} + V_{ij}$ is the quarkonium Hamiltonian. By using Eq.(~\ref{qq})
one gets,
\begin{equation}\label{eq:HPbound}
E_{\rm min}(H_4) \ge 2 \, E_{\rm min}(h_{13}) = 2 \, E_{\rm min}(Q\bar Q) \, ,
\end{equation}
that demonstrates the instability of all-heavy tetraquarks.
The above reasoning on the ground state holds for a single 
color channel. It is observed in explicit computations than 
the mixing of color states does no help much~\cite{Vij09,Via07}. 
The lower bound~(\ref{eq:HPbound}) can even be significantly improved 
if one relates Hamiltonians that are free of center-of-mass motion~\cite{Ric18}.

\section{Color dynamics.}
In the heavy-quark limit, the lowest lying tetraquark configuration
resembles the helium atom~\cite{Eic17}, a factorized system with separate dynamics
for the compact color $\bar 3$ $QQ$ {\em nucleus} and for the
light quarks bound to the stationary color $3$ state, to construct
a $QQ\bar q \bar q$ color singlet. This argument has been mathematically proved and numerically
checked time ago~\cite{Via09}, see the probabilities 
for the axial vector $bb\bar u \bar d$ tetraquark shown in Table II
(note that the $6\bar 6$ probability
in a compact $QQ\bar q\bar q$ tetraquark tends to zero for $M_Q \to \infty$). 

The $\ll{i}{j}$ model of Eq.~(\ref{eq:H4}), with a pairwise potential due to 
color-octet exchange, induces mixing between $\bar33$ and $6\bar6$ states in the 
$QQ-\bar q\bar q$ basis. If one starts from a $\bar33$ 
state with $QQ$ in a spin triplet, and, for instance $\bar q\bar q=\bar u\bar d$ with spin and 
isospin $S=I=0$, then its orbital wave function is mainly made of an $s$-wave in all coordinates. 
It can mix with a color $6\bar 6$ with orbital excitations in the $\vec x$ and $\vec y$ linking $QQ$ 
and $\bar q\bar q$, respectively. A minimal wave function in this sector can be chosen as:
\begin{eqnarray}
 \label{eq:psi6}
\Psi_6 & \propto & \vec x.\vec y\,\exp(-a\,\vec x^{\, 2}-b\,\vec y^{\, 2})~,\nonumber \,\,\,\,\,\,\,\,\, \rm{or}\\
\Psi_6 & \propto & 
 \exp\bigl[-a_{12}\,\vec x^{\, 2}-a_{34}\,\vec y^{\, 2} -\alpha(\vec r_{13}^{\, 2}+\vec r_{24}^{\, 2})
 -\beta(\vec r_{14}^{\, 2}+\vec r_{23}^{\, 2})\bigr]-\{\alpha\leftrightarrow\beta\}~.
\end{eqnarray}
To illustrate the role of color-mixing we use the potential AL1~\cite{Sil94}. 
Its central part is a Coulomb-plus-linear potential. Its spin-spin part is a regularized Breit-Fermi 
interaction, with a smearing parameter that depends on the reduced mass.

\begin{figure}[t]
\vspace*{-1cm}
\centering
\includegraphics[width=.45\columnwidth]{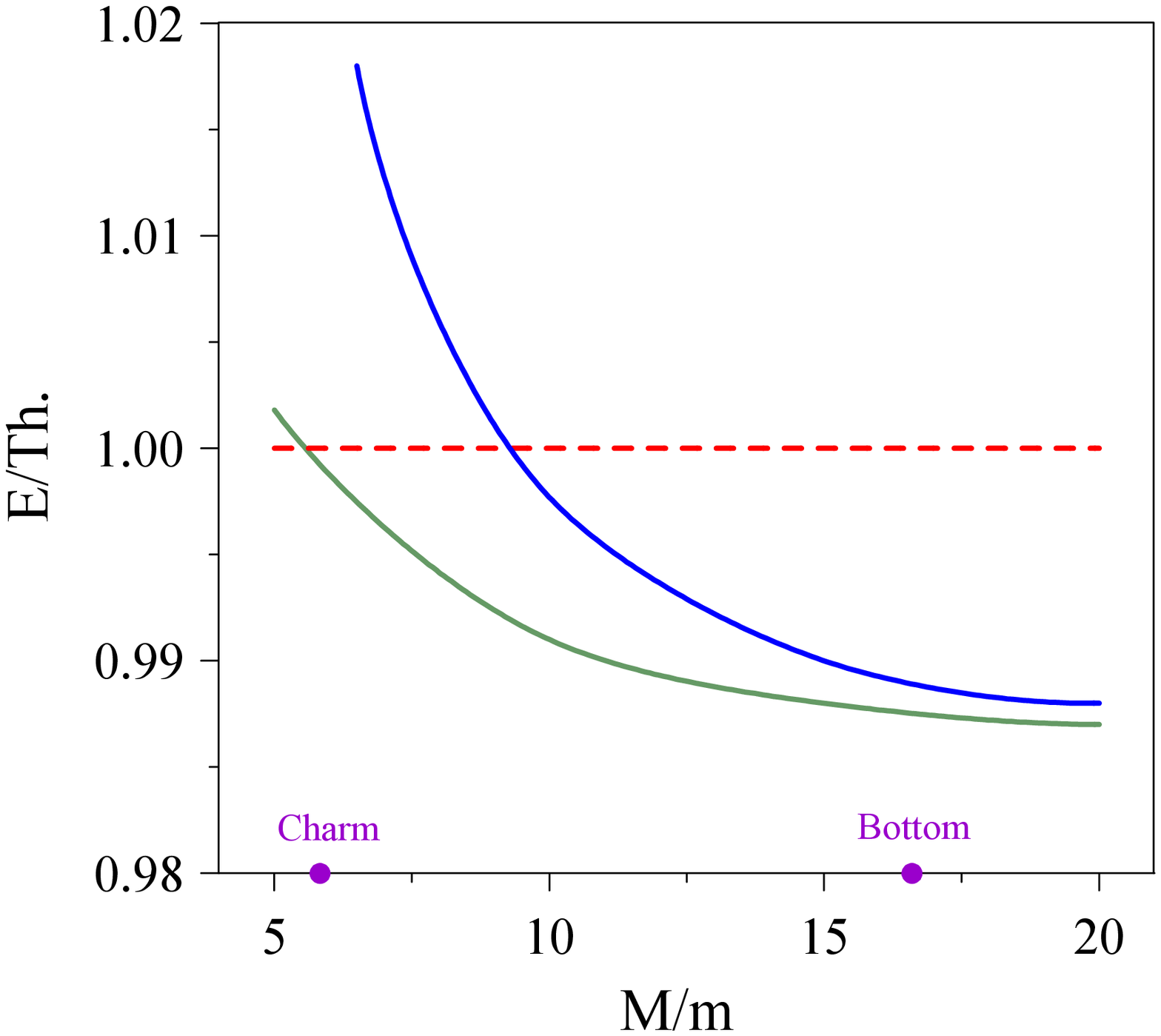}
\includegraphics[width=.45\columnwidth]{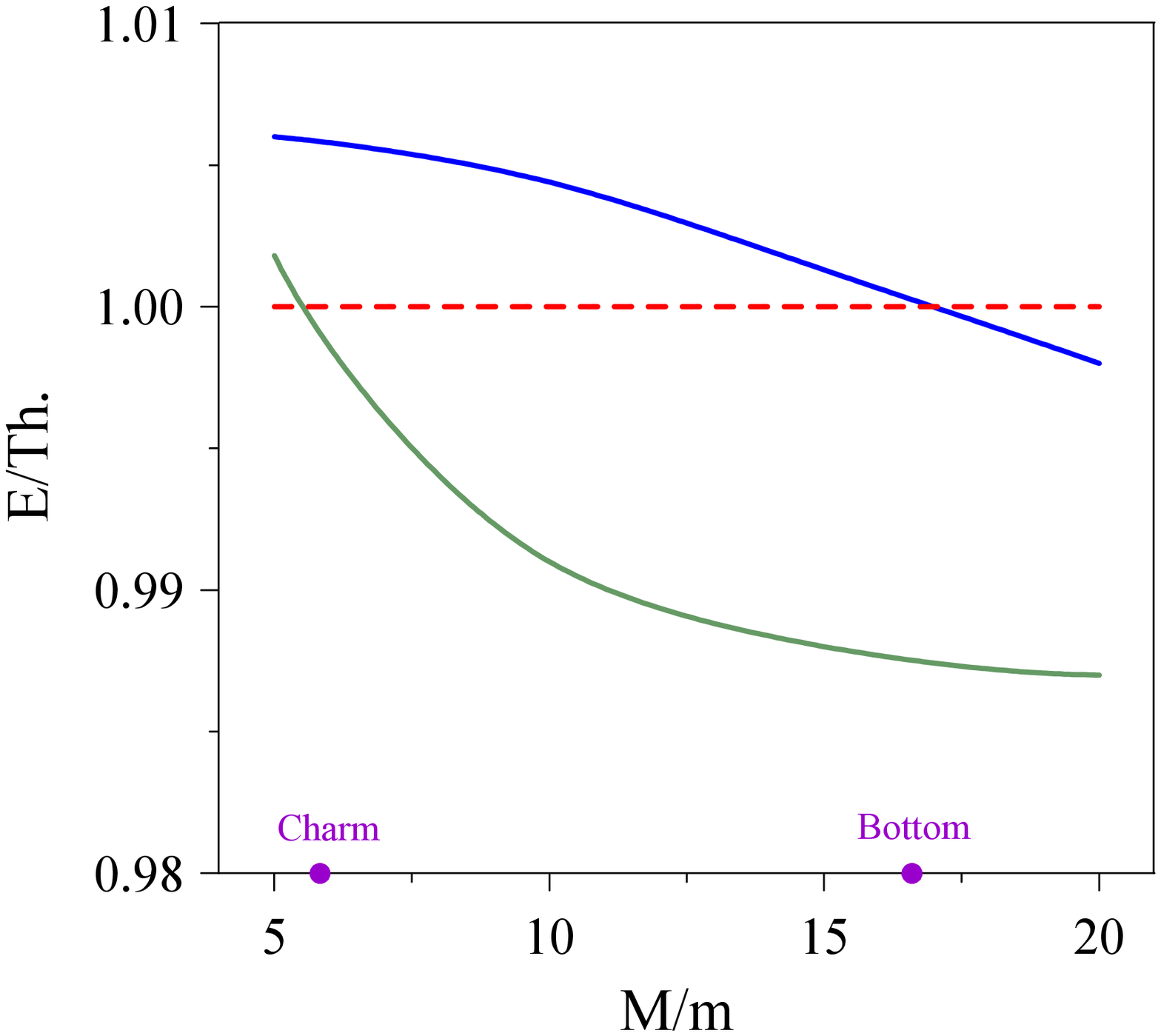}
\vspace*{-4.5cm}
\caption{Left panel: Effect of color-mixing on  the binding of $QQ\bar u\bar d$. 
The tetraquark energy calculated with only the color $\bar33$ configurations (dark curve) and with 
the $6\bar6$ components (light grey curve). Right panel: Effect of the spin-spin interaction of the binding 
of $QQ\bar u\bar d$. The tetraquark energy calculated with (light grey line) and without (dark line) the chromomagnetic term.
The faint dotted solid lines stand for the threshold. Figures are colored online.}
\label{fig:AL1}
\end{figure}

The energy as a function of $M/m$ without and with color-mixing is shown in the left panel of Fig.~\ref{fig:AL1}. 
The ground state of the $QQ\bar u\bar d$, candidate for stability with $J^P=1^+$, has its main component 
with color $\bar33$, and spin $\{1,0\}$ in the $QQ-\bar u\bar d$ basis. The main admixture consists of $6\bar 6$ 
with spin $\{1,0\}$ and an antisymmetric orbital wave function of which (\ref{eq:psi6}) is a prototype, and of 
$6\bar 6$ with spin $\{0,1\}$ with a symmetric orbital wave function.
Note how the diquark hypothesis and color mixing have opposite effects that tend to cancel in the charm sector.

\section{Spin-dependent corrections}\label{se:spin}
In Ref.~\cite{Bal83} it was acknowledged that, within current models,  a pure additive interaction such as
(\ref{eq:H4}) will not bind $cc\bar q\bar q$, on the sole 
basis that this tetraquark configuration benefits from the strong $cc$ chromoelectric attraction 
that is absent in the $Q\bar q+Q\bar q$ threshold. When $qq=ud$, there is in 
addition a favorable chromomagnetic interaction in the tetraquark, while the threshold experiences 
only heavy-light spin-spin interaction, whose strength is suppressed by a factor $m/M$.
 
For illustration, we use the again the potential AL1~\cite{Sil94}. 
The results are shown in the right panel of Fig.~\ref{fig:AL1} for $QQ\bar u\bar d$, as a function of the mass ratio $M/m$.
The system $bb\bar u\bar d$ is barely bound without the spin-spin term, though the mass ratio $m_b/m_q$ is 
very large. It acquires its binding energy of the order of 150\,MeV when the spin-spin is restored. 
The system $cc\bar u\bar d$ is clearly unbound when the spin-spin interaction is switched off. This 
is shown here for the AL1 model, but this is true for any realistic interaction, including an early 
model by Bhaduri {\it et al.}~\cite{Bha81}. The case of $cc\bar u\bar d$ is actually remarkable. 
Here the binding requires both the color mixing of $\bar33$ with $6\bar 6$, and the spin-spin interaction. 
Moreover, the  binding is so tiny that it cannot be obtained with a simple variational method. One needs 
either a fully converged expansion on a basis of correlated Gaussians, or a hyperspherical expansion up to 
a grand orbital momentum $K_{\rm{max}}$ of the order of 12. 
Semay and Silvestre-Brac~\cite{Sil94}, who used the AL1 potential, missed the binding, but their method of 
systematic expansion on the eigenstates of an harmonic oscillator is not very efficient to account 
for the short-range correlations. 
Janc and Rosina~\cite{Jan04} were the first to obtain binding with such potentials, and their calculation 
was checked in Ref.~\cite{Vij09}. The stability of $cc\bar u\bar d$ with $J^P=1^+$ 
is with respect to the nominal $DD^*$ threshold. Depending on its binding energy, 
it decays into $DD\pi$ or $DD\gamma$. The $bb$ analog decays weakly. 

\section{Conclusions}\label{se:concl}
The four-body problem of tetraquarks is rather delicate, especially for systems at the edge of stability.
The analogy with atomic physics is a good guidance to indicate the most favorable configurations. 
However, unlike the positronium molecule, the all-heavy configuration $QQ\bar Q\bar Q$ is not 
stable if one adopts a standard quark model and solve the four-body problem correctly. 
The mixing of the $\bar 3 3$ and $6\bar 6$ color configurations is important, especially for states very near the threshold.  
This mixing occurs by  both  the spin-independent and the spin-dependent parts of the potential. 

Approximations are welcome, especially if they shed some light on the four-body dynamics. The diquark-antidiquark approximation 
is not supported by a rigorous solution of the 4-body problem, but benefits of a stroke of luck, as the erroneous extra attraction 
introduced in the color $\bar 33$ channel is somewhat compensated by the neglect of the coupling to the color $6\bar 6$ channel.  

Finally, $cc\bar u\bar d$ with $J^P=1^+$ is at the edge of binding within current quark models. For this state, all contributions 
should be added, in particular the mixing of states with different color structure, and the four-body problem should be 
solved with extreme accuracy. In comparison, achieving the binding of $bb\bar u\bar d$ looks easier. Still, with a typical quark model, 
the stability of the ground state cannot be reached if spin-effects and color mixing are both neglected. 

\ack
Work funded by Ministerio de Econom\'\i a, Industria y Competitividad
and EU FEDER under Contract No.\ FPA2016-77177 and
by Generalitat Valenciana PrometeoII/2014/066.

\end{document}